\documentclass[prb,twocolumn,amsmath,amssymb,superscriptaddress]{revtex4}
\usepackage{graphicx}
\usepackage{enumitem}
\usepackage{bbold}
\usepackage[colorlinks,bookmarks=false,citecolor=red,linkcolor=blue,urlcolor=blue]{hyperref}
\newcommand{\bea}{\begin{eqnarray}}
\newcommand{\eea}{\end{eqnarray}}
\newcommand{\mbk}{\mathbf{k}}
\newcommand{\bk}{\mathbf{k}}

\newcommand{\dg}{{\dagger}}

\newcommand{\nn}{\nonumber}

\allowdisplaybreaks

\begin{document}

\title{Dirac Hamiltonians for bosonic spectra}
\author{P. Sathish Kumar}
\affiliation{The Institute of Mathematical Sciences, HBNI, C I T Campus, Chennai 600 113, India}
\author{Igor F. Herbut}
\email{igor\_herbut@sfu.ca}
\affiliation{Department of Physics, Simon Fraser University, Burnaby, British Columbia, Canada V5A 1S6}
\author{R. Ganesh}
\email{ganesh@imsc.res.in}
\affiliation{The Institute of Mathematical Sciences, HBNI, C I T Campus, Chennai 600 113, India}

\date{\today}

\begin{abstract}
Dirac materials are of great interest as condensed matter realizations of the Dirac and Weyl equations. In particular, they serve as a starting point for the study of topological phases. This physics has been extensively studied in electronic systems such as graphene, Weyl- and Dirac semi-metals. In contrast, recent studies have highlighted several examples of Dirac-like cones in collective excitation spectra, viz. in phonon, magnon and triplon bands. These cannot be directly related to the Dirac or Weyl equations as they are bosonic in nature with pseudo-unitary band bases. In this article, we show that any Dirac-like equation can be smoothly deformed into a form that is applicable to bosonic bands. The resulting bosonic spectra bear a two-to-one relation to that of the parent Dirac system. Their dispersions inherit several interesting properties including conical band touching points and a gap-opening-role for `mass' terms. The relationship also extends to the band eigenvectors with the bosonic states carrying the same Berry connections as the parent fermionic states. 
The bosonic bands thus inherit topological character as well. If the parent fermionic system has non-trivial topology that leads to mid-gap surface states, the bosonic analogue also hosts surface states that lie within the corresponding band gap.
The proposed bosonic Dirac structure appears in several known models. In materials, it is realized in Ba$_2$CuSi$_2$O$_6$Cl$_2$ and possibly in CoTiO$_3$ as well as in paramagnetic honeycomb ruthenates. Our results allow for a rigorous understanding of Dirac phononic and magnonic systems and enable concrete predictions, e.g., of surface states in magnonic topological insulators and Weyl semi-metals.
\end{abstract}
                                 
\maketitle

\section{Introduction}
Band theory arose from the study of electrons in solid state crystals. Recent developments have revealed a key role for topology here with signatures such as protected surface states and quantized Hall response. A more recent wave of studies has focussed interest on a different class of band structures -- that of collective excitations about ordered phases, such as phonons in crystals or spin waves in ordered magnets\cite{Onose2010,Matsumoto2011,Shindou2013,Hoogdalem2013,Zhang2013,Romhanyi2015}. They also show signatures of non-trivial topology, e.g., in thermal Hall transport. Theoretical investigations have shown that they can also host surface states.
However, these are fundamentally different from electronic systems due to the bosonic nature of the particles involved. This dictates that the band basis must be chosen as a pseudo-unitary set of eigenvectors (defined more precisely below) as opposed to a unitary basis for fermions\cite{Blaizot1986,Bogolubov2009}. On account of this crucial difference, several aspects of fermionic band theory do not carry over to the bosonic problem. The concept of Dirac cones is seemingly one such example that is closely tied to band topology. 
In this article, we show that Dirac cones can indeed be extended to bosonic systems in a precise sense.

Our work is motivated by a flurry of discoveries of cone-like features in bosonic band structures. These include plasmons\cite{Downing2017}, phonons\cite{Huang2012,Yu2015,Yu2016,Gao2016}, photons\cite{Raghu2008,Sakoda2012,Chan2012,He2015,Yi2016}, magnons\cite{Shindou2013,Owerre2016,Fransson2016,Kim2016,Owerre_SciRep_2017} and triplons\cite{Romhanyi2015,McClarty2017,Joshi2017}. There has also been an explosion of interest in `Weyl points' in magnonic band structures\cite{Li2016,Mook2016,Kangkang2017,Li2017,Ying2017,Jian2017}. Do these constitute Dirac systems? Motivated by this question, we identify a class of bosonic Hamiltonians that are adiabatically deformed versions of standard Dirac-like equations. As they retain all key properties of the parent Dirac equation, they can be legitimately labelled as bosonic Dirac systems.

\section{Diagonalizing Bosonic Hamiltonians }
\label{sec.diag}
A fermionic band system is be described by a Hamiltonian, $H =\sum_\mathbf{k} \Psi_{f,\mathbf{k}}^\dagger M_{f,\mathbf{k}} \Psi_{f,\mathbf{k}}$, where $\Psi_{f,\mathbf{k}}$ is a column vector of second quantized annihilation operators. For example, for graphene, we have $\Psi_f = \{  c_{A,\mathbf{k},}, c_{B,\mathbf{k}} \}^T$, where the operators annihilate particles on the $A/B$ sublattices. The Hamiltonian is diagonalized by a transformation 
\bea
\Psi_{f,\mbk} = U_\mbk \Phi_{f,\mbk}; ~~H =\sum_\mathbf{k} \Phi_{f,\mathbf{k}}^\dagger 
M_{f,\mbk}^{diag}
\Phi_{f,\mathbf{k}},
\eea 
where $U_\mathbf{k}$ has the right eigenvectors of $M_{f,\mbk}$ as its columns. This leads to the diagonal form $M_{f,\mbk}^{diag} = U_\mbk^\dagger M_{f,\mbk} U_\mbk $. As $U_\mbk^\dagger U_\mbk = \mathbb{1}$, this can be seen as a unitary transformation that takes us to the `band basis'. The unitary character is of fundamental importance as it preserves the fermionic character of the second quantized operators, i.e., we have $\{ (\Psi_{f,\mbk})_a,(\Psi_{f,\mbk'}^\dagger)_b \} = \{ (\Phi_{f,\mbk})_a,(\Phi_{f,\mbk'}^\dagger)_b \} = \delta_{\mbk,\mbk'} \delta_{a,b}$. This allows us to interpret the new $\Phi$ operators as fermions obeying Fermi-Dirac statistics, e.g., allowing us to identify a Fermi level. This requirement of a unitary transformation holds even in the presence of superconductivity where $\Psi_{f,\mbk}$ mixes annihilation and creation operators.

In a bosonic system, e.g,. that of spin waves in an ordered magnet, we have $H =\sum_\mathbf{k} \Psi_{b,\mathbf{k}}^\dagger M_{b,\mathbf{k}} \Psi_{b,\mathbf{k}}$. 
Generically, $\Psi_{b,\mathbf{k}}$ is a column vector containing both annihilation and creation operators. For example, in the case of magnons in the honeycomb XY ferromagnet, we have 
$\Psi_{b,\mathbf{k}} = (b_{A,\mbk},b_{B,\mbk},b_{A,-\mbk}^\dagger, b_{B,-\mbk}^\dagger)^T$ where $b_{A/B,\mbk}$ represents a Holstein-Primakov operator on the A/B sublattice.
As we have bosonic operators, we have a \textit{commutation} relation given by $[ (\Psi_{b,\mbk})_a,(\Psi_{b,\mbk'}^\dagger)_b ] =  \delta_{\mbk,\mbk'} \mu_{a,b}$. Here, $\mu$ is the commutation matrix that contains $1$'s and $(-1)$'s along the diagonal. For example, in the honeycomb XY ferromagnet, we have $\mu = \mathrm{Diag}\{1,1,-1,-1\}$.

The bosonic nature changes the character of the transformation to the band basis. We require
\bea
\Psi_{b,\mbk} = W_\mbk \Phi_{b,\mbk}; ~~H =\sum_\mathbf{k} \Phi_{b,\mathbf{k}}^\dagger 
M_{b,\mbk}^{diag}
\Phi_{b,\mathbf{k}},
\eea
where $M_{b,\mbk}^{diag} = W_\mbk^\dagger M_{b,\mbk} W_\mbk$. Crucially, in order to preserve the commutation relations, we must have
\bea
W_\mbk^\dagger \mu W_\mbk =W_\mbk \mu W_\mbk^\dagger =  \mu,
\label{eq.pseudounit}
\eea
i.e., $W_\mbk$ must preserve the commutation matrix in order to have $[ (\Phi_{b,\mbk})_a,(\Phi_{b,\mbk'}^\dagger)_b ] =  \delta_{\mbk,\mbk'} \mu_{a,b}$. 
We refer to $W_\mbk$ as a \textit{pseudo-unitary} matrix. The eigenvectors of the band basis here are mutually orthogonal under the pseudo-unitarity condition given by Eq.~\ref{eq.pseudounit}. 

It is perhaps not widely appreciated that the well known properties of Dirac materials follow from unitary diagonalization. For example, the standard definition of Berry connection assumes orthogonality of eigenvectors under unitarity. Likewise, standard proofs for the existence of edge states also assume unitary eigenvectors\cite{Bernevig2013}. These notions cannot be directly adapted to bosonic systems due to the pseudo-unitary nature of bosonic eigenvectors. Below, we demonstrate a deformation procedure that makes this possible.

\section{A fermionic-bosonic deformation }
\label{sec.deform}
We first demonstrate a deformation procedure that takes \textit{any} fermionic Hamiltionian into a bosonic Hamiltonian such that the two spectra have a definite relationship. Consider an $M\times M$ fermionic Hamiltonian matrix, $M_f$. It is diagonalized by a unitary transformation, $U_f$, i.e., $U_f^\dagger M_f U_f = \mathrm{Diag}\{ e_1,e_2,\ldots,e_M\}$, where the $e_i$'s denote the energy eigenvalues.

We now construct a $2M\times 2M$ \textit{bosonic} Hamiltonian,
\bea
\nonumber M_b &=& g_0 \sigma_0 \otimes \mathbb{1}_M + (\sigma_0 + a \cos \theta \sigma_x + a\sin \theta \sigma_y) \otimes M_f \\
&=& \left( 
\begin{array}{cc}
g_0 \mathbb{1}_M + M_f & ae^{-i\theta} M_f \\
ae^{i\theta} M_f & g_0 \mathbb{1}_M + M_f 
\end{array}\right).
\label{eq.Hdeformed}
\eea
Here, $\sigma_0$ and $\mathbb{1}_M$ denote the $2\times 2$ and $M\times M$ identity matrices respectively. $\sigma_{x,y,z}$ are the usual Pauli matrices. We have introduced three real parameters, $g_0$, $a$ and $\theta$. The results below hold for any $\theta$. For practical purposes, we restrict our attention to $a\in [0,1]$. This represents the deformation parameter. At $a=0$, $M_b$ reduces to two independent copies of $M_f$ with an overall shift. For $a>0$, we have a genuine bosonic Hamiltonian that requires a pseudo-unitary transformation.
For physical reasons explained below, the parameter $g_0$ is assumed to be the largest scale in the problem, i.e., $g_0 \gg \vert e_i \vert$ for every $i$.  
Note that the fermionic Hamiltonian, $M_f$, appears in each of the four $M\times M$ blocks of $M_b$. To define $M_b$ as a bosonic Hamiltonian, we must specify the commutation matrix of the operators that it acts upon. We set $\mu = \sigma_z \otimes \mathbb{1}_M$, with $+1$ for the first $M$ diagonal entries and $-1$ for the subsequent $M$ entries.

We require a pseudo-unitary transformation to diagonalize $M_b$. We do this in two steps. We first apply a unitary transformation using 
\bea
\tilde{U} =\sigma_0 \otimes U_f = \left(
\begin{array}{cc}
U_f & 0 \\
0 & U_f 
\end{array}\right).
\eea
As its diagonal blocks contain $U_f$, it diagonalizes the $M_f$ matrix that occurs within each block of $M_b$. We obtain 
\bea
\tilde{U}^\dagger M_b \tilde{U} =
\left(
\begin{array}{ccc|ccc}
\ddots & ~ & ~ & \ddots & ~ & ~ \\
~ & g_0 + e_i  & ~  &  ~ &a e^{-i\theta}e_i & ~  \\
~ & ~ & \ddots & ~& ~ & \ddots \\ \hline
\ddots & ~ & ~ & \ddots & ~ & ~ \\
 ~ &a e^{i\theta}e_i & ~ & ~ & g_0 + e_i  & ~  \\
 ~& ~ & \ddots & ~ & ~ & \ddots
 \end{array}
\right).
\eea
We have a matrix composed of four blocks each of which is diagonal. The eigenvalues of $H_f$ appear in each block in the same order. 
This large matrix is immediately seen to consist of $2\times 2$ blocks by rearrangement. We require an additional transformation that will act within each $2 \times 2$ block to bring about a diagonal form. Such a transformation is carried out by the matrix
\bea
\tilde{P} = \left(
\begin{array}{ccc|ccc}
\ddots & ~ & ~ & \ddots & ~ & ~ \\
~ & \cosh\beta_i & ~  &  ~ &\sinh\beta_i e^{-i\theta} & ~  \\
~ & ~ & \ddots & ~& ~ & \ddots \\ \hline
\ddots & ~ & ~ & \ddots & ~ & ~ \\
 ~ &\sinh\beta_i e^{i\theta} & ~ & ~ & \cosh\beta_i  & ~  \\
 ~& ~ & \ddots & ~ & ~ & \ddots
 \end{array}
\right).
\label{eq.Pmatrix}
\eea
Here, the hyperbolic functions are given by
\bea
\cosh \beta_i = \sqrt{\frac{g_0 + e_i }{2E_i} + \frac{1}{2}} ; ~\sinh\beta_i =
 -\sqrt{\frac{g_0 + e_i }{2E_i} -\frac{1}{2}},
\eea
where 
\bea
E_i =  \sqrt{
g_0^2 + (1-a^2 ) e_i^2 + 
2 g_0 e_i  
}.
\label{eq.boseval}
\eea 
We obtain a diagonal form,
\bea
\tilde{P}^\dagger (\tilde{U}^\dagger M_b \tilde{U} )\tilde{P} = \mathrm{Diag}  \{E_1,\ldots,E_M,E_1\dots\,E_M\}.
\eea
The resulting eigenvalues are given by $E_i$, with a two-fold degeneracy. As each $E_i$ is a function of the corresponding fermionic eigenvalue $e_i$, we have a one-to-two relationship between the fermionic and bosonic spectra. Our assumption of $g_0 \gg \vert e_i\vert$ ensures that the bosonic eigenvalue $E_i$ is positive. As the bosonic Hamiltonian represents collective excitations, a negative energy would be unphysical and indicative of an instability of the underlying ordered phase.

We have performed a transformation on $M_b$ in two steps, with the overall transformation matrix given by a product, $W = \tilde{U} \tilde{P}$. Note that $W$ satisfies the pseudo-unitarity condition with $W^\dagger \mu W = W \mu W^\dagger = \mu$, where $\mu = \sigma_z \otimes \mathbb{1}_M$.
We have expressed the transformation matrix as a product of a unitary matrix, $\tilde{U}$, and a Hermitian matrix, $\tilde{P}$. This can be viewed as an application of `polar decomposition' -- a result in linear algebra that states that any matrix can be decomposed as a product of two matrices, one unitary and one Hermitian. This is analogous to the statement that any complex number can be written as a product of an amplitude and a phase.  

To summarize, the fermionic and bosonic Hamiltonians, $M_f$ and $M_b$, exhibit a close relationship. This is revealed in their spectra: each eigenvalue of $M_f$, $e_i$, determines two eigenvalues of $M_b$ given by $E_i =  \sqrt{
g_0^2 + (1-a^2 ) e_i^2 + 
2 g_0 e_i  
}$. The bosonic energy can be thought of as a smooth function of the fermionic energy, $E(e)$. In particular, $E(e)$ is a monotonic function that increases with increasing $e$. This property follows from our assumption of $g_0 \gg \vert e\vert$.  
The eigenstates of $M_f$ and $M_b$ also exhibit a one-to-two relationship as described in Sec.~\ref{ssec.Berry} below.
This represents a deformation of a generic system from fermionic to bosonic character, with $a=0$ representing the fermionic limit. In this limit, the spectrum of $M_b$ reduces to $\{g_0 + e_i\}$, corresponding to two shifted copies of $M_f$. The transformation matrix becomes purely unitary in this limit, while also preserving the pseudo-unitarity condition. In particular, $\tilde{P}$ reduces to the $2M\times2M$ identity matrix. For $a\neq0$, we have a genuine bosonic problem with $W$ being a pseudo-unitary matrix.

\section{The bosonic Dirac Hamiltonian}
\label{sec.bosDirac}
In the previous section, we have described a deformation of an arbitrary fermionic Hamiltonian into a bosonic problem with one-to-two relationship in the eigenvalue spectrum. Here, we apply this deformation principle to Dirac systems.

As is well known, any Dirac Hamiltonian is written in terms of a `Clifford algebra'\cite{Schweber_book,Okubo1991}. This is a set of $N$ matrices, each of size $M\times M$, such that (a) they each square to identity and (b) they anticommute with one another. Examples include Pauli matrices ($ N=3,M=2$) and Dirac Gamma matrices ($N=5, M=4$). Pauli matrices appear in the Su-Schrieffer Heeger model, graphene, Weyl semi-metals, etc. The Dirac Gamma matrices appear in Dirac semi-metals, quadratic band touching points, etc.

We denote a generic Clifford algebra using $\Gamma_i$, $i=1,\ldots,N$ where $\{ \Gamma_i,\Gamma_j \} = 2\delta_{i,j} \mathbb{1}_M$. We also introduce $\Gamma_{N+1}\equiv \mathbb{1}_M$, the $M\times M$ identity matrix. 
A generic (fermionic) Dirac Hamiltonian is given by
\bea
H_{f,D} = \sum_{i=1}^{N+1} g_i \Gamma_i \equiv g_{N+1} \Gamma_{N+1} + \vec{g}\cdot \vec{\Gamma},
\label{eq.Hferm}
\eea
Here, the $g_{N+1}$ term represents an overall shift. We have grouped the remaining coefficients, $g_i$ with $i=1,\ldots,N$, into an $N$-dimensional vector. This Hamiltonian can be diagonalized by a unitary matrix $U_f$ to give $U_f^\dagger H_{f,D} U_f =
 \mathrm{Diag} \{ g_{N+1} + \vert \vec{g} \vert , \ldots, g_{N+1} - \vert \vec{g} \vert,\ldots  \}$. We have two eigenvalues, $E_{f,D,\pm}=g_{N+1} \pm \vert \vec{g} \vert$, that are $M/2$-fold degenerate. Here, $\vert \vec{g} \vert = \sqrt{\sum_i^{N} g_i^2 }$.

To define a bosonic analogue of the Dirac Hamiltonian, we use the deformation outlined in Sec.~\ref{sec.deform}, 
\bea
\nonumber H_{b,D} &=& g_0 \sigma_0 \otimes \mathbb{1}_M + (\sigma_0 + a \cos \theta \sigma_x + a\sin \theta \sigma_y) \otimes H_{f,D} \\
&=& \left( 
\begin{array}{cc}
g_0 \mathbb{1}_M + H_{f,D} & ae^{-i\theta} H_{f,D} \\
ae^{i\theta} H_{f,D} & g_0 \mathbb{1}_M + H_{f,D} 
\end{array}\right),
\label{eq.Hbosonic}
\eea
where $H_{f,D}$ is as given in Eq.~\ref{eq.Hferm} above.
As described in Sec.~\ref{sec.deform}, we have introduced three free parameters in $a$, $\theta$ and $g_0$. The bosonic spectrum (found using a pseudo-unitary transformation) of this Hamiltonian can be deduced from Eq.~\ref{eq.boseval}, 
\bea
E_{b,D,\pm}=
 \sqrt{ A \pm 2 B \vert \vec{g} \vert}, 
\label{eq.eigB}
\eea
where $A = g_0^2 + (1-a^2 ) \{ g_{N+1}^2 + \vert \vec{g} \vert^2 \}+ 2g_0 g_{N+1}$ and $B = g_0+ (1-a^2) g_{N+1}$. We note that there are two distinct bosonic eigenvalues, each of which is $M$-fold degenerate.

\section{Features inherited in the bosonic Dirac problem}
We have discussed a relationship between a fermionic Dirac system, described by the Hamiltonian in Eq.~\ref{eq.Hferm}, and a bosonic problem with the Hamiltonian given in Eq.~\ref{eq.Hbosonic}. We now enumerate the characteristic features of the former that are inherited by the latter.  

\subsection{Dirac cone structure}

The fermionic and bosonic Dirac spectra both consist of two bands. We first note that, in both spectra, the bands touch if and only if $\vec{g} = 0$. Secondly, we recapitulate a general property that was pointed out in Sec.~\ref{sec.deform} above: the bosonic energy is a smooth, monotonic function of the fermionic energy. From these two statements, we conclude that: (i) a band gap in the fermionic spectrum implies a corresponding gap in the bosonic spectrum, and (ii) if the fermionic spectrum is gapless with a Dirac cone, the same holds for the bosonic system as well. To see this clearly, we restrict our attention to the immediate vicinity of a Dirac point where $\vec{g}=0$. Considering $g_{N+1}$ and all components of $\vec{g}$ to be small, we may keep them to linear order in a Taylor expansion. We find $E_{b,D,\pm}\approx
 g_0 +g_{N+1}\pm \vert \vec{g} \vert$, with a clear Dirac cone structure. 
 
In condensed matter realizations, the components of $\vec{g}$ are typically functions of momenta that transform as representations of a rotation group. The Dirac cone feature arises when they are proportional to momenta, $\mathbf{p}$ (or more precisely, deviations from a particular point in momentum space). This leads to the dispersion $E_{f,D,\pm}\sim g_{N+1}\pm v_F\vert \mathbf{p} \vert $. Likewise, in the bosonic problem, if the components of $\vec{g}$ are proportional to momenta, we find $E_{b,D,\pm} \sim g_0 + g_{N+1} \pm v_F\vert \mathbf{p} \vert$. Note that the bosonic Dirac cone is centred at $g_0 + g_{N+1}$ whereas the fermionic Dirac cone is centred at $g_{N+1}$. In particular, if $g_{N+1}=0$, the fermionic system has negative eigenvalues whereas the bosonic system always has positive energies (assuming  $g_0\gg \vec{g}$). This is consistent with the stability requirement that forbids collective modes with negative energies.

\subsection{Role of mass terms}
 
In the fermionic problem, if a Dirac cone is realized using only a subset of $g_i$'s in $H_f$, a gap can be opened by introducing one of the remaining $\Gamma$ matrices with a non-zero coefficient (i.e., a coefficient that is not zero at the location of the Dirac point). Such a term is called a `mass term'\cite{Herbut2009,Ryu2009}. If such a mass term of strength $\Delta$ is introduced, we have $E_{f,D,\pm} \sim g_{N+1} \pm \vert \Delta \vert + \mathcal{O}(\vert \vec{g}\vert^2)$, with the two bands separated by a gap of $2 \vert \Delta\vert$. This has a direct analogue in the bosonic problem. In Eq.~\ref{eq.Hbosonic}, this new term simply enters as an added component in $H_{f,D}$. This leads to the bosonic dispersion $E_{b,D,\pm} = C \pm D \vert \Delta \vert + \mathcal{O}(\vert \vec{g} \vert^2  )$, where $C = \sqrt{(g_0 + g_{N+1})^2 - a^2 g_{N+1}^2}$ and $D = \{ g_0 + (1-a^2) g_{N+1}\}/C$. In particular, if $g_{N+1}$ is small, we have $E_{b,D,\pm} = g_0 + g_{N+1} \pm \vert \Delta \vert + \mathcal{O}(\vert \Delta \vert/g_0,g_{N+1}^2,\vert \vec{g} \vert^2  )$.

\subsection{Berry connection}
\label{ssec.Berry}
A band gap induced by mass terms in a Dirac system may have topological character. 
This can be determined using invariants that are built from `Berry connections' associated with the change in a band eigenvector as the Hamiltonian parameters are tuned. Here, we show that the bosonic bands inherit the Berry connections of the parent fermionic bands. In the process, we discuss a one-to-two relationship between the \textit{eigenstates} of the fermionic and bosonic Dirac systems. 

The eigenstates of the fermionic Dirac Hamiltonian can be thought of as columns of a unitary matrix. It is this matrix, $U_f$, that diagonalizes the Hamiltonian. We now denote $U_f$ in terms of its sub-blocks,
\begin{eqnarray}
U_f = \left(\begin{array}{c|c}
u_1 & u_2 
\end{array}\right),
\end{eqnarray}
where the $u_1$ contains the first $M/2$ columns and $u_2$ contains the last $M/2$ columns. Both $u_1$ and $u_2$ represent $M\times M/2$ blocks. Note that in any Clifford algebra, the size of the matrices, M, is necessarily even\cite{Schweber_book}. This allows us to split the diagonalization matrix into two parts, each with $M/2$ columns. 
For the bosonic Hamiltonian described by $M_b$, the eigenstates are found as columns of the pseudo-unitary matrix $W=\tilde{U}\tilde{P}$ defined in Sec.~\ref{sec.deform}. For the bosonic Dirac system of Eq.~\ref{eq.Hbosonic}, the $\tilde{P}$ matrix given in Eq.~\ref{eq.Pmatrix} takes a particularly simple form. This leads to
\begin{eqnarray}
W 
=\left(
\begin{array}{c|c|c|c}
u_1 C_+ & u_2 C_- & u_1 S_+ e^{-i\theta} & u_2 S_- e^{-i\theta} \\ \hline
u_1 S_+ e^{i\theta} & u_2 S_- e^{i\theta}  & u_1 C_+ & u_2 C_-
\end{array}\right),
\label{eq.Wexp}
\end{eqnarray}
with each block representing an $M \times M/2$ matrix. Here, $C_\pm$ and $S_\pm$ denote hyperbolic cosine and sine functions, given by
\bea
\nonumber C_\pm &=& \sqrt{\frac{g_0 + g_{N+1}\pm \vert \vec{g} \vert }{2E_{b,D\pm}} + \frac{1}{2}} ; \\
S_\pm &=&
 -\sqrt{\frac{g_0 + g_{N+1}\pm \vert \vec{g} \vert }{2E_{b,D,\pm}} -\frac{1}{2}},
\eea
where the eigenvalues $E_{b,D,\pm}$ are defined in Eq.~\ref{eq.eigB} above. 
We can now see the relationship between the $M$ fermionic and $2M$ bosonic eigenstates, i.e., between the columns of $U_f$ and $W$. 
Each column of $U_f$ is part of two columns of $W$. For example, the first column of ${U_f}$ (i.e., the first column of $u_1$) appears in the first and $(M+1)^{th}$ columns of $W$. 

The relationship also extends to the Berry connection, a property of eigenstates that is central to band topology. We first note that the fermionic Hamiltonian is written in terms of the parameters $g_i$, $i=1,\ldots,N+1$. We now consider the $\ell^{\mathrm{th}}$ eigenstate of the Hamiltonian, contained in the $\ell^{\mathrm{th}}$ column of $U_f$. Its Berry connection is given by  
\begin{eqnarray}
A_{i,\ell}^f = \langle \ell \vert\frac{\partial}{\partial g_i}\vert \ell \rangle =  \sum_{j=1}^M ({U_f}^\dagger)_{\ell,j} \frac{\partial}{\partial g_i} {U_f}_{j,\ell}.
\end{eqnarray}
Here, $\ell$ represents a band while $i$ represents a parameter in the Hamiltonian. The form of this Berry connection is tied to the unitary nature of eigenvector matrix, i.e., 
$U_f^{-1} = U_f^\dagger$. For concreteness, we now consider $\ell \leq M/2$ so that the $\ell^{th}$ column of $U_f$ consists of entries in $u_1$. We have
\begin{eqnarray}
A_{i,\ell}^f = \sum_{j=1}^{M} \left[ 
(u_1^\dagger)_{\ell,j} \frac{\partial}{\partial g_\alpha} (u_1)_{j,\ell}
\right].
\label{eq.Aferm}
\end{eqnarray}
We now consider the bosonic Hamiltonian which is determined by the same parameters, $g_i$, $i=1,\ldots, N+1$. Note that $g_0$, $\theta$ and $a$ are taken to be fixed parameters. The Berry connection of the $\ell^{\mathrm{th}}$ column of $W$ (assuming $\ell \leq M/2$), is given by,
\begin{eqnarray}
A_{i,\ell}^b = \sum_{j=1}^{2M} (W^{-1})_{\ell,j} \frac{\partial}{\partial g_i} W_{j,\ell},
\end{eqnarray}
Here, we have $W^{-1}$ rather than $W^\dagger$, as $W$ is not unitary. The inverse is obtained from the pseudo-unitary condition, giving $W^{-1} = \mu W^\dagger \mu$. We obtain
\begin{eqnarray}
\nonumber A_{i,\ell}^b = \sum_{j=1}^{M} \Big[
&~& C_+ (u_1^\dagger)_{\ell,j} \frac{\partial}{\partial g_i}\big\{ (u_1)_{j,\ell} C_+ \big\} -\\
&~&  S_+ e^{-i\theta}(u_1^\dagger)_{\ell,j} \frac{\partial}{\partial g_i}\big\{ (u_1)_{j,\ell} S_+ e^{i\theta} \big\}~\Big].~~
\end{eqnarray}
With a few simplifications using hyperbolic trigonometric identities, we separate out terms that constitute the fermionic Berry connection, $A_{i,\ell}^f$. Remarkably, all other terms vanish so that
\bea
A_{i,\ell}^b = A_{i,\ell}^f .
\eea
The Berry connection of the bosonic band is the same as that of its fermionic parent. The argument carries through for all bosonic eigenstates. Notably, as pointed out earlier, there are two bosonic eigenstates that derive from a single fermionic state. Both have the same Berry connection as the parent fermionic state.
It follows that a one-to-two relationship exists in terms of the Chern number as well. The bosonic bands have the same topological character as the parent fermionic bands.

\subsection{Surface states}

A fermionic Dirac system may have a topologically non-trivial gap. This leads to protected surface states when translational symmetry is broken, e.g., upon introducing an edge or a spatially modulated mass term. The resulting states have energies that lie within the erstwhile gap. Their wavefunctions are localized on the surface, decaying exponentially into the bulk. We now show that this property is also inherited by a bosonic Dirac system.

Let us first consider a fermionic Dirac Hamiltonian with unbroken translational symmetry and with a topologically non-trivial gap. In this case, the Dirac Hamiltonian will take the form given in Eq.~\ref{eq.Hferm} with the coefficients, $g_i$, being constants or functions of momenta. The dispersion will have a well-defined gap, say from $(-e_{g},e_g)$. In the deformed bosonic problem, we will have an analogous gap from $(E_{g,-},E_{g,+})$, where $E_{g,\pm} = \sqrt{g_0^2 + (1-a^2) e_g^2 \pm 2g_0 e_g}$. 

We now introduce a spatial modulation in the fermionic problem. If the modulation is smooth, the Hamiltonian takes the form
\begin{eqnarray}
\hat{H}_{f,\mathrm{edge}}  = \sum_{i=1}^{N+1} g_i (x,{\partial_ x}) \Gamma_i,
\end{eqnarray} 
where the coefficients, $g_i$, are functions of space (e.g., spatially modulated mass terms) or of spatial derivatives (e.g., a smooth edge with $k_x \rightarrow i \partial_x$). The Hamiltonian should now be thought of as an operator acting on the space of differentiable functions.
The non-trivial topology guarantees the existence of eigenstates that are (a) localized along edges and (b) have energies within the gap. We consider one such state, $\psi_f(\mathbf{r})$,
\bea
\hat{H}_{f,\mathrm{edge}} \psi_f(\mathbf{r}) = \Big(\sum_{i=1}^{N+1} g_i (x,{\partial_ x}) \Gamma_i\Big) \psi_f(\mathbf{r})  = \tilde{e}  \psi_f(\mathbf{r}),
\eea
where $-e_g < \tilde{e} < e_g$. The wavefunction $\psi_f(\mathbf{r})$ satisfies suitable boundary conditions, e.g., it may vanish as $x\rightarrow 0$ and decay exponentially for large $x$.

The corresponding bosonic Hamiltonian is given by
\begin{eqnarray}
\nonumber \hat{H}_b  &=& g_0 \sigma_0 \times \mathbb{1}
+ \sum_i^{N+1} g_i (x,{\partial_ x}) \times \\
&~&(\sigma_0 + a \cos \theta \sigma_x + a\sin\theta \sigma_y) \otimes \Gamma_i.
\end{eqnarray} 
We will construct solutions for this Hamiltonian below. As the Hamiltonian is bosonic, we require an eigenstate of $(\mu \hat{H}_b)$ rather than $\hat{H}_b$ itself, where the commutation matrix $\mu=\sigma_z \otimes \mathbb{1}_\mathbf{r}$, where $\mathbb{1}_\mathbf{r}$ is the identity operator. This is a consequence of pseudo-unitary diagonalization. As $W^\dagger M_b W = M_{diag} \implies W^{-1} \mu M_b W = \mu M_{diag}$, the diagonal matrix is obtained as a similarity transformation on $\mu M_b$, rather than on $M_b$ itself. 

We construct two bosonic eigenstates, in analogy with the translationally symmetric case discussed above.
\bea
\psi_b (\mathbf{r}) \!=\! \left(
\begin{array}{cc}
\psi_f(\mathbf{r}) \cosh \beta \\
\psi_f(\mathbf{r}) \sinh \beta e^{i\theta} 
\end{array}\right)\!,\!
\psi'_b (\mathbf{r}) \!=\! \left(
\begin{array}{cc}
\psi_f(\mathbf{r}) \sinh \beta e^{-i\theta} \\
\psi_f(\mathbf{r}) \cosh \beta  
\end{array}\right)\!\!,~
\eea
where 
\bea 
\cosh \beta = \sqrt{\frac{g_0 + \tilde{e}}{2\tilde{E}} + \frac{1}{2}},~\sinh\beta =
 -\sqrt{\frac{g_0 + \tilde{e}}{2\tilde{E}} -\frac{1}{2}},
 \eea
with $\tilde{E}= \sqrt{g_0^2 + (1-a^2) \tilde{e}^2 + 2g_0 \tilde{e}}$.
These two states are indeed eigenfunctions of $(\mu \hat{H}_b)$. This can be seen as 
\bea
\nonumber &~& \mu \hat{H}_b \psi_b (\mathbf{r}) = \\
\nonumber &~&\left(
\begin{array}{cc}
g_0 + \hat{H}_{f,edge} & ae^{-i\theta} \hat{H}_{f,edge}\\
-ae^{i\theta} \hat{H}_{f,edge} & -g_0 - \hat{H}_{f,edge}
\end{array}\right)\left(
\begin{array}{cc}
\psi_f(\mathbf{r}) \cosh \beta \\
\psi_f(\mathbf{r}) \sinh \beta e^{i\theta} 
\end{array}\right) \\
&~&=\tilde{E} \left(
\begin{array}{cc}
\psi_f(\mathbf{r}) \cosh \beta \\
\psi_f(\mathbf{r}) \sinh \beta e^{i\theta} 
\end{array}\right),
\eea
where we have used hyperbolic trigonometric identities. In the same manner, we also have $ (\mu \hat{H}_b) \psi'_b (\mathbf{r}) = -\tilde{E} \psi'_b (\mathbf{r})$. 

We have constructed two bosonic eigenstates using a fermionic surface state, $\psi_f(\mathbf{r})$, as a building block. They are degenerate and their eigenvalue, $\tilde{E}$, is determined by $\tilde{e}$,  the eigenvalue of the fermionic surface state. Thus, the one-to-two relationship extends to surface states as well. 
From the expression for $\tilde{E}$ above, we see that $E_{g,-} < \tilde{E} <E_{g,+}$, i.e., the bosonic edge states lie within the band gap of the translationally symmetric bosonic Dirac problem.
We note the spatial dependence $\psi_b(\mathbf{r})$ and $\psi'_b(\mathbf{r})$ enters via $\psi_f(\mathbf{r})$. Thus, both $\psi_b(\mathbf{r})$ and $\psi'_b(\mathbf{r})$ satisfy the same boundary conditions as the fermionic surface state -- they are also surface states.  

We have shown that if the fermionic Dirac problem has surface states, so does the deformed bosonic problem. We have used a long wavelength approach here, taking the Hamiltonian to be a differential operator. This approach is suitable for `soft' edges which take the form of slowly varying potentials. In the case of a hard edge, it would be more appropriate to use a microscopic tight binding model with open boundary conditions. Here again, given a fermionic tight binding problem, we can deform it into a bosonic problem using the procedure outlined in Sec.~\ref{sec.deform}. Note that the arguments of Sec.~\ref{sec.deform} are for an arbitrary $M\times M$ fermionic Hamiltonian, readily applicable to a tight binding setting. A band gap in the translationally symmetric problem implies a corresponding band gap in the deformed bosonic problem as well. Likewise, in-gap surface states in the fermionic problem with open boundary conditions are inherited by the bosonic problem. 

\subsection{Squaring to find eigenvalues}
Any fermionic Dirac Hamiltonian has a remarkable property which allows for its eigenvalues to be found by mere inspection. This stems from a simple diagonal form that emerges from squaring the Hamiltonian.
To see this explicitly, we consider the generic $M \times M$ fermionic Dirac Hamiltonian of Eq.~\ref{eq.Hferm}. It is diagonalized by a unitary matrix $U_f$, with $ U_f U_f^\dagger = \mathbb{1}_M$ and $U_f^\dagger H_{f,D} U_f = H_{f,\mathrm{diag}}$ being diagonal. It is immediately seen that this matrix also diagonalizes $(H_{f,D})^2$ since $U_f^\dagger H_{f,D}^2 U_f = (U_f^\dagger H_{f,D} U_f)^2 = H_{f,\mathrm{diag}}^2$. Here, we have inserted an identity matrix in the middle, in the form of $U_f U_f^\dagger$. This reveals an important relationship: the eigenvalues of $H_{f,D}$ are simply the square-roots of those of $H_{f,D}^2$. While this relationship holds true for any fermionic Hamiltonian, it is particularly useful in the case of Dirac Hamiltonians. To see this, we modify Eq.~\ref{eq.Hferm} to define $\tilde{H}_{f,D} \equiv H_{f,D} - g_{N+1} \Gamma_{N+1} = \vec{g}\cdot \vec{\Gamma}$. As this is composed of Clifford algebra matrices, we have
\bea
\tilde{H}_{f,D}^2 = \sum_{i,j=1}^N g_ i g_j \{ \Gamma_i,\Gamma_j\} =  \vert \vec{g} \vert ^2 \mathbb{1}_M.
\eea
Here, we have used the anticommutation properties of Clifford algebra matrices. We see that $(\tilde{H}_{f,D})^2$ is a simple diagonal matrix whose entries (and eigenvalues) are all $\vert \vec{g} \vert^2$. We conclude that the eigenvalues of $\tilde{H}_{f,D}$ are $\pm \vert \vec{g} \vert$. We further deduce that the eigenvalues of $H_{f,D}$ are $g_{N+1} \pm \vert \vec{g} \vert$. 

This `squaring trick' is a fundamental property of the Dirac equation that underlies all of its interesting features. Although it may not be appreciated at first sight, it relies crucially on the unitary nature of the diagonalization transformation, i.e., it requires $U_f U_f^\dagger = \mathbb{1}_M$. Naively, this precludes its use in bosonic systems which require pseudo-unitary transformations. Nevertheless, we now demonstrate an analogous property for $H_{b,D}$, the bosonic Dirac Hamiltonian of Eq.~\ref{eq.Hbosonic}. 

Let us suppose the eigenvalues of $H_{b,D}$, given by $(E_1,\ldots,E_{2M})$, are to be determined. The diagonalizing transformation is carried out by a $2M \times 2M$ matrix, $W$, that satisfies $W \mu W^\dagger = \mu$ and $W^\dagger H_{b,D} W = \mathrm{Diag}\{E_1,\ldots,E_{2M}\}$. Here, $\mu=\sigma_z \otimes \mathbb{1}_M$ is the commutation matrix as described in Sec.~\ref{sec.diag}. Unlike fermionic systems, the matrix $W$ does not diagonalize $H_{b,D}^2$. Rather, it diagonalizes $(H_{b,D} \mu H_{b,D})$, 
\bea
\nonumber &~& W^\dagger (H_{b,D} \mu H_{b,D}) W = (W^\dagger H_{b,D} W) \mu (W^\dagger H_{b,D} W) \\
 &~&~~~~~~~= \{E_1^2,\ldots,E_{2M}^2,-E_1^2,\ldots,-E_{2M}^2 \}.~~
\eea
Here, we have replaced $\mu$ in the middle with $W \mu W^\dagger $, using the pseudo-unitary condition. We see that the (bosonic) eigenvalues of $(H_{b,D}\mu H_{b,D})$ are closely related to those of $H_{b,D}$. This relationship is true for any Hamiltonian in the place of $H_{b,D}$. However, it becomes particularly useful in the bosonic Dirac Hamiltonian given in Eq.~\ref{eq.Hbosonic}. It leads to a simple form for $(H_{b,D}\mu H_{b,D})$, whose eigenvalues can be found `by inspection'. The arguments presented above can then be used to find the (bosonic) eigenvalues of $H_{b,D}$.

To show this, we consider the explicit form of $H_{b,D}$ in Eq.~\ref{eq.Hbosonic}. After a few simplifications, we obtain
\bea
H_{b,D}  \mu  H_{b,D} =
\left(
\begin{array}{cc} 
A \mathbb{1}_M + 2 B \vec{g} \cdot \vec{\Gamma} &  0_M \\
 0_M & -A \mathbb{1}_M - 2 B \vec{g} \cdot \vec{\Gamma}
\end{array}
\right),~~
\label{eq.HmuH}
\eea
where the coefficients $A$ and $B$ are as defined below Eq.~\ref{eq.eigB}.
At this stage in the fermionic problem, we had arrived at a diagonal matrix. Here, we have arrived at a block diagonal form with zeros in the off-diagonal blocks. Nevertheless, the eigenvalues of Eq.~\ref{eq.HmuH} can be found by inspection. To see this, we first note that the diagonal blocks are (apart from a shift) proportional to the previously discussed fermionic matrix, $\tilde{H}_{f,D} = \vec{g}\cdot \vec{\Gamma}$. Its (unitary) eigenvalues are known from the arguments above. 
We define a transformation matrix, $\tilde{W}$, given by $\tilde{W} = \sigma_0 \otimes U_f$, where $U_f$ is unitary and diagonalizes $\tilde{H}_{f,D}$. The matrix $\tilde{W}$ is clearly unitary, however it also satisfies the pseudo-unitary condition with $\tilde{W} \mu \tilde{W}^\dagger = \mu$. We can immediately see that $\tilde{W}$ diagonalizes the Hamiltonian in Eq.~\ref{eq.HmuH} above. The resulting eigenvalues are $A\pm 2 B \vert \vec{g} \vert$ and $-A\mp 2 B \vert \vec{g} \vert$ . This allows us to deduce the eigenvalues of $H_b$ which take the same form as Eq.~\ref{eq.eigB} above.

To summarize the arguments in this section, there exists a squaring trick for the bosonic Dirac Hamiltonian, $H_{b,D}$. This stems from a simple relationship between the bosonic eigenvalues of $H$ and $(H\mu H)$, where $H$ is any bosonic Hamiltonian and $\mu$ is the commutation matrix. In bosonic Dirac Hamiltonians, $(H \mu H)$ takes a simple form whose eigenvalues can be found immediately. They are found using a unitary transformation that also satisfies the pseudo-unitary condition. This has a deep underlying reason that can be understood from the discussion in Sec.~\ref{sec.deform}. The bosonic Dirac Hamiltonian is diagonalized in two steps: a unitary transformation followed by a pseudo-unitary transformation. The same overall transformation also diagonalizes $(H_{b,D} \mu H_{b,D})$. However, in the case of $(H_{b,D} \mu H_{b,D})$, the unitary transformation alone suffices. It converts the Hamiltonian into a specific diagonal form that is not altered by the subsequent pseudo-unitary transformation. The bosonic Dirac Hamiltonian can be viewed as a specific form that is designed to achieve this. 

\section{Examples in model systems}
We demonstrate the bosonic Dirac structure in two simple magnetic systems that show a Dirac cone feature. We subsequently discuss model systems from literature that contain this structure.

\subsection{XY ferromagnet on the honeycomb lattice}
We first consider the honeycomb XY ferromagnet, described by the Hamiltonian
\begin{eqnarray}
\nonumber H_{XY} = -J\sum_{i} \sum_{\delta} \left[ S_{i,A}^x S_{i+\delta,B}^x + S_{i,A}^y S_{i+\delta,B}^y \right],
\end{eqnarray}
where $i$ runs over all unit cells of the honeycomb lattice, with each unit cell containing two sites, labelled $A$ and $B$. The three nearest neighbours of a given A-sublattice site are denoted by $(i+\delta,B)$, the B site of the unit cell at $(i+\delta)$. Here, $\delta$ takes three possible values. The ground state of this system exhibits long range order due to spontaneous symmetry breaking, e.g., with ferromagnetic moment along the X direction. The excitations about this ordered state are spin waves or magnons, with the Hamiltonian,
\begin{eqnarray}
H =   JS\sum_{\mathbf{k}}{}^{'} \Phi_\mathbf{k}^\dagger H_{\mathbf{k}}^b \Phi_\mathbf{k} + \mathrm{const.}
\label{eq.Hamiltonianbosons}
\end{eqnarray} 
The primed summation signifies that if $\mathbf{k}$ is included in the sum, $-\mathbf{k}$ must be excluded. The vector of operators $\Phi_\mathbf{k}$ and the Hamitonian matrix are given by
\begin{eqnarray}
\nonumber
 \Phi_\mathbf{k} &=& \left(  \begin{array}{cccc}
a_{\mathbf{k}} &
b_{\mathbf{k}} &
a_{-\mathbf{k}}^\dagger &
b_{-\mathbf{k}}^\dagger
\end{array}
 \right)^T, \\ 
 H_{\mathbf{k}}^b &=&
\left( 
\begin{array}{cccc}
 3&-\epsilon_\mathbf{k} &0 & \epsilon_\mathbf{k}\\ 
  -{\epsilon}^{*}_\mathbf{k}&3 &{\epsilon}^{*}_\mathbf{k} &0 \\
  0 & \epsilon_\mathbf{k} &3 & -\epsilon_\mathbf{k}\\  
  {\epsilon}^{*}_\mathbf{k}&0 &-{\epsilon}^{*}_\mathbf{k}   & 3
 \end{array}\right) \\
 \nonumber &=& 3 \{ \sigma_0 \otimes \sigma_0 \} -\mathrm{Re}(\epsilon_\mbk) \{ (\sigma_0 - \sigma_x)\otimes
 \sigma_x\}  \\
 &~& ~~~~~~~~~~~~~~~ +\mathrm{Im}(\epsilon_\mbk)\{(\sigma_0 - \sigma_x)\otimes
 \sigma_y \}. 
 \label{eq.phi_H_honeycombXY}
 \end{eqnarray} 
Here, the operators are given by the Holstein Primakoff prescription with $a_i^\dagger$ and $b_i^\dagger$ creating spin excitations on the A and B sites of the unit cell labelled by $i$.
These operators are then expressed in momentum space, e.g., $a_\mathbf{k} = \sum_{i\in A} a_i e^{i\mathbf{k}\cdot \mathbf{r}_i}$. We have defined $\epsilon_\mathbf{k}=\frac{1}{2}\sum_\delta e^{i\mathbf{k}\cdot \mathbf{\delta}}$.

This Hamiltonian conforms to our bosonic Dirac prescription. It has the form of Eq.~\ref{eq.Hbosonic} with the Clifford algebra matrices, $\Gamma_i$'s, taken to be Pauli matrices. 
The parameters in the bosonic deformation take the values, ($g_0=3$, $a=1$, $\theta=\pi$).  The parent fermionic Dirac Hamiltonian here is $H_{f,D} =  -\mathrm{Re}(\epsilon_\mbk) \sigma_x +  \mathrm{Im}(\epsilon_\mbk) \sigma_y$, resembling graphene or more precisely, spinless fermions hopping on a honeycomb lattice. As is well known, this problem possesses Dirac cones at the K point with the dispersion taking the form $E_f\sim \pm \vert \mathbf{q} \vert$ in its vicinity. Here, $\mathbf{q}$ represents deviation from the K point. The bosonic spin wave Hamiltonian inherits this Dirac cone structure. From the results deduced in Sec.~\ref{sec.deform}, we see that it has the dispersion, $E_b \sim JS\sqrt{9 \pm 6\vert \mathbf{q} \vert^2} \approx JS \{3 \pm \vert \mathbf{q}\vert\}$.

This model and its cone-like dispersion have been discussed in Ref.~\onlinecite{Owerre2016} as an example of a magnonic Dirac point. Our analysis places this identification on firm footing, as a smooth deformation of spinless fermions on a honeycomb lattice. Ref.~\onlinecite{Owerre2016} goes on to invoke a next-nearest neighbour Dzyaloshinskii-Moriya coupling as a perturbation that opens a gap. Interestingly, this perturbation is not a bonafide mass term even though it opens a gap. 
In line with the arguments above, a genuine mass term must have the form, $H_{mass,\mbk} \sim (\sigma_0 - \sigma_x)\otimes \sigma_z$. Such a term arises from next-nearest neighbour $y-y$ couplings,
\begin{eqnarray}
\nonumber H_{mass} &=& J\Delta \sum_{i\in A} \sum_{\eta} \left[ S_{i,A}^y S_{i+\eta,A}^y - S_{i,B}^y S_{i+\eta,B}^y\right]  \\
\nonumber &=& J\Delta S
\sum_{\mathbf{k}}{}^{'} \zeta_\mathbf{k} \left[ a_{\mathbf{k}}^\dagger a_{\mathbf{k}} - a_{\mathbf{k}}^\dagger a_{-\mathbf{k}}^\dagger - a_{-\mathbf{k}} a_{\mathbf{k}} + a_{-\mathbf{k}} a_{-\mathbf{k}}^\dagger \right. \\
\nonumber &~&~~~~~~~~\left. - b_{\mathbf{k}}^\dagger b_{\mathbf{k}} + b_{\mathbf{k}}^\dagger b_{-\mathbf{k}}^\dagger + b_{-\mathbf{k}} b_{\mathbf{k}} - b_{-\mathbf{k}} b_{-\mathbf{k}}^\dagger 
  \right] \\
  &\Rightarrow&   \Delta \sum_{\mathbf{k}}{}^{'} \zeta_\mathbf{k}  (\sigma_0-\sigma_x)\otimes \sigma_z.
\end{eqnarray}
The index $\eta$ runs over the six next-nearest neighbours in the honeycomb lattice.
We use $\zeta_\mathbf{k} = \sum_\eta e^{i\mathbf{k}\cdot \eta}$. This term opens a gap in the spectrum with $E_{b}= JS \sqrt{9 \pm 6 \sqrt{\vert \mathbf{q}\vert^2 + \Delta^2} } \approx JS\{3 \pm 2\Delta + \mathcal{O}(q^2,\Delta^2) \}$.

\subsection{Interlayer honeycomb valence bond solid}
\label{ssec.ilvbs}
We next consider a $S=1/2$ honeycomb bilayer magnet with dominant interlayer antiferromagnetic couplings, denoted by $J_\perp$. The intra-layer nearest-neighbour couplings are denoted by $J_1$. A similar Hamiltonian is realized in ${\mathrm{Bi}}_{3}{\mathrm{Mn}}_{4}{\mathrm{O}}_{12}({\mathrm{NO}}_{3})$\cite{Alaei2017}. 
With $J_\perp \gg J_1$, the ground state is a valence bond solid with a singlet formed on each $J_\perp$ bond\cite{Ganesh2011,Vishwanath2004}. 
The excitations about the valence bond solid correspond to breaking a singlet and allowing the resulting triplet to move. Such `triplon' excitations are known to be bosonic, well-described by the bond operator prescription of Sachdev and Bhatt\cite{Sachdev1990}. In this approach, a mean-field singlet amplitude, $\bar{s}$, and a chemical potential, $\mu$, are introduced to preserve the physical spin Hilbert space. The resulting `triplon' Hamiltonian is given by\cite{Ganesh2011}
\bea
\label{hcmfthamiltonian}
\nn H = \mathrm{const}.
+\sum_{\bk,u}{}^{'}
\psi_{\bk,u}^\dg M_\mbk \psi_{\bk,u}, \phantom{ab}
\eea
Here, $u=x,y,z$ denotes the three `flavours' of triplons that can be created on a dimer.
 The operator $\psi_{\mbk,u}$ and the Hamiltonian matrix $M_\bk$ are given by
\bea
\label{Eq.hcmftpsiM}
\psi_{\bk,u}=\left(\begin{array}{c}
t_{\bk,A,u} \\ t_{\bk,B,u} \\ t_{-\bk,A,u}^\dg \\  t_{-\bk,B,u}^\dg 
\end{array}\right), \phantom{ab}
M_\bk =
\left(\begin{array}{cccc}
C & \beta_\bk & 0 & \beta_\bk \\
\beta_\bk^* & C & \beta_\bk^* & 0 \\
0 & \beta_\bk & C & \beta_\bk \\
\beta_\bk^* & 0 & \beta_\bk^* & C
\end{array}\right),
\eea
where $C=(3J_\perp/4-\mu)$ and $\beta_\bk= \frac{1}{4}\bar{s}^2 \epsilon_\bk$, where $\epsilon_\bk$ has been defined below Eq.~\ref{eq.phi_H_honeycombXY}. The $A/B$ sublattices denote the sublattice where the interlayer dimer sits. 

This Hamiltonian is of the bosonic Dirac form given in Eq.~\ref{eq.Hbosonic} with $(g_0 = C, a= 1, \theta =0)$. The parent fermionic Dirac Hamiltonian, once again, corresponds to spinless fermions on the honeycomb lattice. The triplon spectrum inherits Dirac cones at the $K$ points, as shown in Ref.~\onlinecite{Ganesh2011}. 

\subsection{Other models with bosonic Dirac structure}
In the literature on bosonic systems with non-trivial topology, we find two other realizations of the bosonic Dirac structure proposed here. The results known in these contexts can now be re-interpreted in terms of Dirac Hamiltonians. 

(i) A triplonic analogue of the Su-Schrieffer-Heeger model has been proposed by Joshi and Schnyder\cite{Joshi2017}. Their triplon Hamiltonian can be thought of as a weakly deformed Dirac system. It can be brought into the bosonic Dirac form of Eq.~\ref{eq.Hbosonic} by (a) tuning the inter-dimer exchange to zero ($K=0$) and (b) introducing the magnetic field into $H_2(k)$ so as to make $\vec{x}$ and $\vec{d}$ identical (in their notation). This demonstrates that this ladder system represents a deformation of the Su-Schrieffer-Heeger model. As a result, it has a topological invariant in the form of a winding number. Indeed, the authors have evaluated the winding number using $W^{-1}$ instead of $W^\dagger$, as in Sec.~\ref{ssec.Berry} above. The one-to-two relation to the Su-Schrieffer-Heeger model also explains the occurrence of edge states in the topological regime.

(ii) Joshi and Schnyder have also presented a bosonic version of a $\mathbb{Z}_2$ topological insulator\cite{Joshi2019}. This involves a honeycomb bilayer antiferromagnet with a valence bond solid ground state, similar to Sec.~\ref{ssec.ilvbs} above. The triplon excitations form two bands separated a topological band gap. This system is a straightforward realization of the bosonic Dirac Hamiltonian of Eq.~\ref{eq.Hbosonic}. The parent fermionic Hamiltonian here is the Kane-Mele model\cite{KaneMele2005} that represents a time-reversal-invariant topological insulator. While the authors have indeed pointed out the similarity to the Kane-Mele model, our analysis establishes a rigorous relationship between the two models.

\section{Examples in materials}
\label{sec.materials}
In many magnetic systems, the Hamiltonians that describe spin excitations are known to high accuracy.
We discuss examples from literature where the Hamiltonians resemble the bosonic Dirac form. 

(i) A direct realization occurs in Ba$_2$CuSi$_2$O$_6$Cl$_2$\cite{Nawa2019}, a dimerized magnet with triplon excitations. The triplon Hamiltonian is of the bosonic Dirac form in Eq.~\ref{eq.Hbosonic}, with $(a=1,\theta=0)$. In order to understand the topology of the triplon band structure, the authors in Ref.~\onlinecite{Nawa2019} disregard the pairing terms and arrive at a fermionic Dirac Hamiltonian. Our analysis shows that the full Hamiltonian, without discarding any terms, realizes a bosonic Dirac system.

(ii) The three dimensional magnet CoTiO$_3$ is composed of stacked honeycomb layers\cite{Yuan2019}. The spins order order ferromagnetically within each layer and antiferromagnetically in the inter-layer direction. Neutron scattering measurements show that the spin wave spectrum has Dirac-like cones. The spectrum has been fit to $XXZ$-type models with varying anisotropy strengths\cite{Yuan2019}. One of the models that has been discussed has XY couplings between nearest and next-nearest neighbours. This represents a precise realization of the bosonic Dirac structure. The other models can be viewed as distortions from the XY limit. It is conceivable that they are connected by smooth deformations to the bosonic Dirac problem.

(iii) A spin-orbital valence bond solid state has been proposed in Mott insulators with two holes in the t$_{2g}$ level\cite{Khaliullin2013}. Each site contains an electronic spin ($S=1$) and an orbital angular momentum ($L=1$) that are antiferromagnetically coupled by spin-orbit coupling. This leads to a dimer-like $J=0$ state at each site with triplon-like excitations\cite{Anisimov2019}. In a honeycomb geometry, realized in Li$_2$RuO$_3$ and Ag$_3$LiRu$_2$O$_6$, this closely resembles the interlayer valence bond solid discussed in Sec.~\ref{ssec.ilvbs} above, but with Kitaev-like inter-dimer couplings. 
Ref.~\onlinecite{Anisimov2019} discusses topology and edge states by neglecting pairing terms. However, this is not necessary as the full Hamiltonian\cite{Anisimov2019thesis} is of the bosonic Dirac form given in Eq.~\ref{eq.Hbosonic}. The parent Hamiltonian describes spinless fermions on the honeycomb lattice with anisotropic hopping.

(iv) An interesting situation emerges in SrCu$_2$(BO$_3$)$_2$, well known as a realization of the Shastry-Sutherland model. Being a valence bond solid, it has triplon excitations that have been described by quantitatively-accurate models. Ref.~\onlinecite{Romhanyi2011} gives the explicit Hamiltonian at zero momentum, as is appropriate for electron spin resonance measurements. 
Ref.~\onlinecite{McClarty2017} gives explicit expressions for the Hamiltonian at all momenta.
The Hamiltonian does not conform to the bosonic Dirac form given in Eq.~\ref{eq.Hbosonic}. Nevertheless, it represents a deformed bosonic Hamiltonian as given in Eq.~\ref{eq.Hdeformed}, as long as (a) the external magnetic field is turned off and (b) the small intra-dimer anisotropy couplings ($J_{xy}$ and $J_{zz}$ in the notation of Ref.~\onlinecite{McClarty2017}) are ignored. With these assumptions, the system can be viewed as a smooth deformation of a parent fermionic Hamiltonian. The parent Hamiltonian can be expressed in an elegant form 
using matrices that form a spin-1 representation of $SU(2)$\cite{Romhanyi2015}. This `spin-1 Dirac cone' structure was identified in an earlier study by neglecting pairing terms in the Hamiltonian\cite{Romhanyi2015}. 
With the perspective presented in this article, it is no longer necessary to neglect pairing terms. The full bosonic Hamiltonian can be obtained as a smooth deformation of a fermionic spin-1 Hamiltonian.

\section{Discussion}
We have proposed a precise notion of Dirac Hamiltonians in bosonic spectra with pseudo-unitary band bases. Our analysis shows that several, if not all, interesting properties of the Dirac equation can be realized in collective mode spectra. This paves the way for identifying a new class of Dirac materials aside from electronic systems\cite{Wehling2014} in states with long range order, particularly in magnets with magnon/triplon excitations. Our work builds upon efforts to extend the notions of topological invariants to bosonic systems\cite{Raghu2008,Shindou2013,Joshi2017,Kondo2019,Kondo2019b}. In a broader sense, we have demonstrated a scheme to extend the notion of Dirac systems by adiabatic deformation. This point of view may be useful in various systems where band structures are obtained by solving generalized eigenvalue problems.

Our analysis concerns bosonic systems that require pseudo-unitary diagonalization. It does not apply to certain magnonic systems where unitary band bases can be used. This is the case in ordered magnets with residual $U(1)$ symmetry, e.g., in Heisenberg magnets with collinear order. This symmetry leads to preservation of magnon number, thereby forbidding  pairing terms ($\sim a^\dagger a^\dagger$). In such cases, fermion-like unitary diagonalization suffices. This class includes several materials such as CrB$_3$\cite{Pershoguba2018}, Cu$_3$TeO$_6$\cite{Yao2018,Bao2018}, Cu(1,3-bdc)\cite{Lee2015} and pyrochlore ferromagnets\cite{Mook2016,Su2017}. Nevertheless, these systems should be considered as the exception rather than as the rule. Pairing terms and pseudo-unitary diagonalization are an unavoidable feature in a wider class of models and materials. Our bosonic deformation procedure offers precise insights in their presence.

Magnonic Weyl semi-metals have received tremendous interest recently. They are three-dimensional magnets with cone-like band touching points. Explicit Hamiltonians have been constructed in two contexts: breathing pyrochlore antiferromagnets\cite{Li2016,Jian2017} as well as stacked Kagome antiferromagnets\cite{Owerre2018}. In both cases, the Hamiltonian is very close to the bosonic Dirac form. It differs only in additional phases that emerge in the pairing terms. We believe their physics can be explained by deforming away from the bosonic Dirac limit. This is an interesting future direction that can potentially classify all possible magnonic Weyl semi-metals.

\acknowledgments
IFH acknowledges support from NSERC of Canada.
\bibliographystyle{apsrev4-1} 
\bibliography{bosonic_clifford}

\end{document}